

%
%


\def\famname{
 \textfont0=\textrm \scriptfont0=\scriptrm
 \scriptscriptfont0=\sscriptrm
 \textfont1=\textmi \scriptfont1=\scriptmi
 \scriptscriptfont1=\sscriptmi
 \textfont2=\textsy \scriptfont2=\scriptsy \scriptscriptfont2=\sscriptsy
 \textfont3=\textex \scriptfont3=\textex \scriptscriptfont3=\textex
 \textfont4=\textbf \scriptfont4=\scriptbf \scriptscriptfont4=\sscriptbf
 \skewchar\textmi='177 \skewchar\scriptmi='177
 \skewchar\sscriptmi='177
 \skewchar\textsy='60 \skewchar\scriptsy='60
 \skewchar\sscriptsy='60
 \def\rm{\fam0 \textrm} \def\bf{\fam4 \textbf}}
\def\sca#1{scaled\magstep#1} \def\scah{scaled\magstephalf} 
\def\twelvepoint{
 \font\textrm=cmr12 \font\scriptrm=cmr8 \font\sscriptrm=cmr6
 \font\textmi=cmmi12 \font\scriptmi=cmmi8 \font\sscriptmi=cmmi6 
 \font\textsy=cmsy10 \sca1 \font\scriptsy=cmsy8
 \font\sscriptsy=cmsy6
 \font\textex=cmex10 \sca1
 \font\textbf=cmbx12 \font\scriptbf=cmbx8 \font\sscriptbf=cmbx6
 \font\it=cmti12
 \font\sectfont=cmbx12 \sca1
 \font\sectmath=cmmib10 \sca2
 \font\sectsymb=cmbsy10 \sca2
 \font\refrm=cmr10 \scah \font\refit=cmti10 \scah
 \font\refbf=cmbx10 \scah
 \def\twelverm{\textrm} \def\twelveit{\it} \def\twelvebf{\textbf}
 \famname \textrm 
 \advance\voffset by .06in \advance\hoffset by .28in
 \normalbaselineskip=17.5pt plus 1pt \baselineskip=\normalbaselineskip
 \parindent=21pt
 \setbox\strutbox=\hbox{\vrule height10.5pt depth4pt width0pt}}


\catcode`@=11

{\catcode`\'=\active \def'{{}^\bgroup\prim@s}}

\def\screwcount{\alloc@0\count\countdef\insc@unt}   
\def\screwdimen{\alloc@1\dimen\dimendef\insc@unt} 
\def\screwbox{\alloc@4\box\chardef\insc@unt}

\catcode`@=12


\overfullrule=0pt			
\vsize=9in \hsize=6in
\lineskip=0pt				
\abovedisplayskip=1.2em plus.3em minus.9em 
\belowdisplayskip=1.2em plus.3em minus.9em	
\abovedisplayshortskip=0em plus.3em	
\belowdisplayshortskip=.7em plus.3em minus.4em	
\parindent=21pt
\setbox\strutbox=\hbox{\vrule height10.5pt depth4pt width0pt}
\def\makefootline{\baselineskip=30pt \line{\the\footline}}
\footline={\ifnum\count0=1 \hfil \else\hss\twelverm\folio\hss \fi}
\pageno=1


\def\put(#1,#2)#3{\screwdimen\unit  \unit=1in
	\vbox to0pt{\kern-#2\unit\hbox{\kern#1\unit
	\vbox{#3}}\vss}\nointerlineskip}


\def\\{\hfil\break}
\def\newpage{\vfill\eject}
\def\center{\leftskip=0pt plus 1fill \rightskip=\leftskip \parindent=0pt
 \def\textindent##1{\par\hangindent21pt\footrm\noindent\hskip21pt
 \llap{##1\enspace}\ignorespaces}\par}
\def\unnarrower{\leftskip=0pt \rightskip=\leftskip}


\def\vol#1 {{\refbf#1} }		 


\def\NP #1 {{\refit Nucl. Phys.} {\refbf B{#1}} }
\def\PL #1 {{\refit Phys. Lett.} {\refbf{#1}} }
\def\PR #1 {{\refit Phys. Rev. Lett.} {\refbf{#1}} }
\def\PRD #1 {{\refit Phys. Rev.} {\refbf D{#1}} }


\hyphenation{pre-print}
\hyphenation{quan-ti-za-tion}

%
%


\def\oonoo#1#2#3{\vbox{\ialign{##\crcr
	\hfil\hfil\hfil{$#3{#1}$}\hfil\crcr\noalign{\kern1pt\nointerlineskip}
	$#3{#2}$\crcr}}}
\def\oon#1#2{\mathchoice{\oonoo{#1}{#2}{\displaystyle}}
	{\oonoo{#1}{#2}{\textstyle}}{\oonoo{#1}{#2}{\scriptstyle}}
	{\oonoo{#1}{#2}{\scriptscriptstyle}}}
\def\dt#1{\oon{\hbox{\bf .}}{#1}}  
\def\ddt#1{\oon{\hbox{\bf .\kern-1pt.}}#1}    
\def\slap#1#2{\setbox0=\hbox{$#1{#2}$}
	#2\kern-\wd0{\hfuzz=1pt\hbox to\wd0{\hfil$#1{/}$\hfil}}}
\def\sla#1{\mathpalette\slap{#1}}                
\def\bop#1{\setbox0=\hbox{$#1M$}\mkern1.5mu
	\lower.02\ht0\vbox{\hrule height0pt depth.06\ht0
	\hbox{\vrule width.06\ht0 height.9\ht0 \kern.9\ht0
	\vrule width.06\ht0}\hrule height.06\ht0}\mkern1.5mu}
\def\bo{{\mathpalette\bop{}}}                        
\def~{\widetilde} 
\mathcode`\*="702A                  
\def\in{\relax\ifmmode\mathchar"3232\else{\refit in\/}\fi} 
\def\f#1#2{{\textstyle{#1\over#2}}}	   
\def\half{{\textstyle{1\over{\raise.1ex\hbox{$\scriptstyle{2}$}}}}}

\def\Gamma{\mathchar"0100}
\def\Delta{\mathchar"0101}
\def\Theta{\mathchar"0102}
\def\Lambda{\mathchar"0103}
\def\Xi{\mathchar"0104}
\def\Pi{\mathchar"0105}
\def\Sigma{\mathchar"0106}
\def\Upsilon{\mathchar"0107}
\def\Phi{\mathchar"0108}
\def\Psi{\mathchar"0109}
\def\Omega{\mathchar"010A}

\catcode128=13 \def €{\"A}                 
\catcode129=13 \def {\AA}                 
\catcode130=13 \def '{\c}           	   
\catcode131=13 \def ƒ{\'E}                   
\catcode132=13 \def "{\~N}                   
\catcode133=13 \def …{\"O}                 
\catcode134=13 \def †{\"U}                  
\catcode135=13 \def ‡{\'a}                  
\catcode136=13 \def ˆ{\`a}                   
\catcode137=13 \def ‰{\^a}                 
\catcode138=13 \def Š{\"a}                 
\catcode139=13 \def ‹{\~a}                   
\catcode140=13 \def Œ{\alpha}            
\catcode141=13 \def {\chi}                
\catcode142=13 \def Ž{\'e}                   
\catcode143=13 \def {\`e}                    
\catcode144=13 \def {\^e}                  
\catcode145=13 \def '{\"e}                
\catcode146=13 \def '{\'\i}                 
\catcode147=13 \def "{\`\i}                  
\catcode148=13 \def "{\^\i}                
\catcode149=13 \def •{\"\i}                
\catcode150=13 \def –{\~n}                  
\catcode151=13 \def —{\'o}                 
\catcode152=13 \def ˜{\`o}                  
\catcode153=13 \def ™{\^o}                
\catcode154=13 \def š{\"o}                 
\catcode155=13 \def ›{\~o}                  
\catcode156=13 \def œ{\'u}                  
\catcode157=13 \def {\`u}                  
\catcode158=13 \def ž{\^u}                
\catcode159=13 \def Ÿ{\"u}                
\catcode160=13 \def  {\tau}               
\catcode161=13 \mathchardef ¡="2203     
\catcode162=13 \def ¢{\oplus}           
\catcode163=13 \def £{\relax\ifmmode\to\else\itemize\fi} 
\catcode164=13 \def ¤{\subset}	  
\catcode165=13 \def ¥{\infty}           
\catcode166=13 \def ¦{\mp}                
\catcode167=13 \def §{\sigma}           
\catcode168=13 \def ¨{\rho}               
\catcode169=13 \def ©{\gamma}         
\catcode170=13 \def ª{\leftrightarrow} 
\catcode171=13 \def «{\relax\ifmmode\acute\else\expandafter\'\fi}
\catcode172=13 \def ¬{\relax\ifmmode\expandafter\ddt\else\expandafter\"\fi}
\catcode173=13 \def ­{\equiv}            
\catcode174=13 \def ®{\approx}          
\catcode175=13 \def ¯{\Omega}          
\catcode176=13 \def °{\otimes}          
\catcode177=13 \def ±{\ne}                 
\catcode178=13 \def ²{\le}                   
\catcode179=13 \def ³{\ge}                  
\catcode180=13 \def ´{\upsilon}          
\catcode181=13 \def µ{\mu}                
\catcode182=13 \def ¶{\delta}             
\catcode183=13 \def ·{\epsilon}          
\catcode184=13 \def ¸{\Pi}                  
\catcode185=13 \def ¹{\pi}                  
\catcode186=13 \def º{\beta}               
\catcode187=13 \def »{\partial}           
\catcode188=13 \def ¼{\nobreak\ }       
\catcode189=13 \def ½{\zeta}               
\catcode190=13 \def ¾{\sim}                 
\catcode191=13 \def ¿{\omega}           
\catcode192=13 \def À{\dt}                     
\catcode193=13 \def Á{\gets}                
\catcode194=13 \def Â{\lambda}           
\catcode195=13 \def Ã{\nu}                   
\catcode196=13 \def Ä{\phi}                  
\catcode197=13 \def Å{\xi}                     
\catcode198=13 \def Æ{\psi}                  
\catcode199=13 \def Ç{\int}                    
\catcode200=13 \def È{\oint}                 
\catcode201=13 \def É{\relax\ifmmode\cdot\else\vol\fi}    
\catcode202=13 \def Ê{\relax\ifmmode\,\else\thinspace\fi}
\catcode203=13 \def Ë{\`A}                      
\catcode204=13 \def Ì{\~A}                      
\catcode205=13 \def Í{\~O}                      
\catcode206=13 \def Î{\Theta}              
\catcode207=13 \def Ï{\theta}               
\catcode208=13 \def Ð{\relax\ifmmode\bar\else\expandafter\=\fi}
\catcode209=13 \def Ñ{\overline}             
\catcode210=13 \def Ò{\langle}               
\catcode211=13 \def Ó{\relax\ifmmode\{\else\ital\fi}      
\catcode212=13 \def Ô{\rangle}               
\catcode213=13 \def Õ{\}}                        
\catcode214=13 \def Ö{\sla}                      
\catcode215=13 \def ×{\relax\ifmmode\check\else\expandafter\v\fi}
\catcode216=13 \def Ø{\"y}                     
\catcode217=13 \def Ù{\"Y}  		    
\catcode218=13 \def Ú{\Leftarrow}       
\catcode219=13 \def Û{\Leftrightarrow}       
\catcode220=13 \def Ü{\relax\ifmmode\Rightarrow\else\sect\fi}
\catcode221=13 \def Ý{\sum}                  
\catcode222=13 \def Þ{\prod}                 
\catcode223=13 \def ß{\widehat}              
\catcode224=13 \def à{\pm}                     
\catcode225=13 \def á{\nabla}                
\catcode226=13 \def â{\quad}                 
\catcode227=13 \def ã{\in}               	
\catcode228=13 \def ä{\star}      	      
\catcode229=13 \def å{\sqrt}                   
\catcode230=13 \def æ{\^E}			
\catcode231=13 \def ç{\Upsilon}              
\catcode232=13 \def è{\"E}    	   	 
\catcode233=13 \def é{\`E}               	  
\catcode234=13 \def ê{\Sigma}                
\catcode235=13 \def ë{\Delta}                 
\catcode236=13 \def ì{\Phi}                     
\catcode237=13 \def í{\`I}        		   
\catcode238=13 \def î{\iota}        	     
\catcode239=13 \def ï{\Psi}                     
\catcode240=13 \def ð{\times}                  
\catcode241=13 \def ñ{\Lambda}             
\catcode242=13 \def ò{\cdots}                
\catcode243=13 \def ó{\^U}			
\catcode244=13 \def ô{\`U}    	              
\catcode245=13 \def õ{\bo}                       
\catcode246=13 \def ö{\relax\ifmmode\hat\else\expandafter\^\fi}
\catcode247=13 \def÷{\relax\ifmmode\tilde\else\expandafter\~\fi}
\catcode248=13 \def ø{\ll}                         
\catcode249=13 \def ù{\gg}                       
\catcode250=13 \def ú{\eta}                      
\catcode251=13 \def û{\kappa}                  
\catcode252=13 \def ü{\half}     		 
\catcode253=13 \def ý{\Gamma} 		
\catcode254=13 \def þ{\Xi}   			
\catcode255=13 \def ÿ{\relax\ifmmode{}^{\dagger}{}\else\dag\fi}


\def\ital#1Õ{{\it#1\/}}	     
\def\un#1{\relax\ifmmode\underline#1\else $\underline{\hbox{#1}}$
	\relax\fi}

\def\roonoo#1#2#3{\vbox{\ialign{##\crcr
	\hfil{$#3{#1}$}\hfil\crcr\noalign{\kern1pt\nointerlineskip}
	$#3{#2}$\crcr}}}
\def\roon#1#2{\mathchoice{\roonoo{#1}{#2}{\displaystyle}}
	{\roonoo{#1}{#2}{\textstyle}}{\roonoo{#1}{#2}{\scriptstyle}}
	{\roonoo{#1}{#2}{\scriptscriptstyle}}}
\def\rdt#1{\roon{\hbox{\bf .}}{#1}}  
\def\tdt#1{\oon{\hbox{\bf .\kern-1pt.\kern-1pt.}}#1}   
\def\({\eqno(}
\def\li{\openup1\jot \eqalignno}


\def\õ#1{
	\screwcount\num
	\num=1
	\screwdimen\downsy
	\downsy=-1.5ex
	\mkern-3.5mu
	õ
	\loop
	\ifnum\num<#1
	\llap{\raise\num\downsy\hbox{$õ$}}
	\advance\num by1
	\repeat}
\def\upõ#1#2{\screwcount\numup
	\numup=#1
	\advance\numup by-1
	\screwdimen\upsy
	\upsy=.75ex
	\mkern3.5mu
	\raise\numup\upsy\hbox{$#2$}}



\newcount\marknumber	\marknumber=1
\newcount\countdp \newcount\countwd \newcount\countht 

%
%
\ifx\pdfoutput\undefined
\def\rgboo#1{}
\input epsf

\def\postscript#1{\special{" #1}}		
\postscript{
	/bd {bind def} bind def
	/fsd {findfont exch scalefont def} bd
	/sms {setfont moveto show} bd
	/ms {moveto show} bd
	/pdfmark where		
	{pop} {userdict /pdfmark /cleartomark load put} ifelse
	[ /PageMode /UseOutlines		
	/DOCVIEW pdfmark}
\def\bookmark#1#2{\postscript{		
	[ /Dest /MyDest\the\marknumber /View [ /XYZ null null null ] /DEST pdfmark
	[ /Title (#2) /Count #1 /Dest /MyDest\the\marknumber /OUT pdfmark}%
	\advance\marknumber by1}
\def\pdfklink#1#2{%
	\hskip-.25em\setbox0=\hbox{#1}%
		\countdp=\dp0 \countwd=\wd0 \countht=\ht0%
		\divide\countdp by65536 \divide\countwd by65536%
			\divide\countht by65536%
		\advance\countdp by1 \advance\countwd by1%
			\advance\countht by1%
		\def\linkdp{\the\countdp} \def\linkwd{\the\countwd}%
			\def\linkht{\the\countht}%
	\postscript{
		[ /Rect [ -1.5 -\linkdp.0 0\linkwd.0 0\linkht.5 ] 
		/Border [ 0 0 0 ]
		/Action << /Subtype /URI /URI (#2) >>
		/Subtype /Link
		/ANN pdfmark}{\rgb{1 0 0}{#1}}}
%
%
\else
\def\rgboo#1{\pdfliteral{#1 rg #1 RG}}

\pdfcatalog{/PageMode /UseOutlines}		
\def\bookmark#1#2{
	\pdfdest num \marknumber xyz
	\pdfoutline goto num \marknumber count #1 {#2}
	\advance\marknumber by1}
\def\pdfklink#1#2{%
	\noindent\pdfstartlink user
		{/Subtype /Link
		/Border [ 0 0 0 ]
		/A << /S /URI /URI (#2) >>}{\rgb{1 0 0}{#1}}%
	\pdfendlink}
\fi

\def\rgbo#1#2{\rgboo{#1}#2\rgboo{0 0 0}}
\def\rgb#1#2{\mark{#1}\rgbo{#1}{#2}\mark{0 0 0}}
\def\pdflink#1{\pdfklink{#1}{#1}}
\def\xxxlink#1{\pdfklink{[arXiv:#1]}{http://arXiv.org/abs/#1}}

\catcode`@=11

\def\wlog#1{}	


\def\makeheadline{\vbox to\z@{\vskip-36.5\p@
	\line{\vbox to8.5\p@{}\the\headline%
	\ifnum\pageno=\z@\rgboo{0 0 0}\else\rgboo{\topmark}\fi%
	}\vss}\nointerlineskip}
\headline={
	\ifnum\pageno=\z@
		\hfil
	\else
		\ifnum\pageno<\z@
			\ifodd\pageno
				\tenrm\romannumeral-\pageno\hfil\lefthead\hfil
			\else
				\tenrm\hfil\righthead\hfil\romannumeral-\pageno
			\fi
		\else
			\ifodd\pageno
				\tenrm\hfil\righthead\hfil\number\pageno
			\else
				\tenrm\number\pageno\hfil\lefthead\hfil
			\fi
		\fi
	\fi}

\catcode`@=12

\def\righthead{\hfil} \def\lefthead{\hfil}
\nopagenumbers


\def\chrulefill{\rgb{1 0 0}{\hrulefill}}
\def\cdotfill{\rgb{1 0 0}{\dotfill}}
\newcount\area	\area=1
\newcount\cross	\cross=1
\def\volume#1\par{\newpage\noindent{\biggest{\rgb{1 .5 0}{#1}}}
	\par\nobreak\bigskip\medskip\area=0}
\def\chapskip{\par\ifnum\area=0\bigskip\medskip\goodbreak
	\else\newpage\fi}
\def\chapy#1{\area=1\cross=0
	\xdef\lefthead{\rgbo{1 0 .5}{#1}}\vbox{\biggerer\offinterlineskip
	\line{\chrulefill¼\hphantom{\lefthead}\chrulefill}
	\line{\chrulefill¼\lefthead\chrulefill}}\par\nobreak\medskip}
\def\chap#1\par{\chapskip\bookmark3{#1}\chapy{#1}}
\def\sectskip{\par\ifnum\cross=0\bigskip\medskip\goodbreak
	\else\newpage\fi}
\def\secty#1{\cross=1
	\xdef\righthead{\rgbo{1 0 1}{#1}}\vbox{\bigger\offinterlineskip
	\line{\cdotfill¼\hphantom{\righthead}\cdotfill}
	\line{\cdotfill¼\righthead\cdotfill}}\par\nobreak\medskip}
\def\sect#1 #2\par{\sectskip\bookmark{#1}{#2}\secty{#2}}
\def\subsectskip{\par\ifdim\lastskip<\medskipamount
	\bigskip\medskip\goodbreak\else\nobreak\fi}
\def\subsecty#1{\noindent{\sectfont{\rgbo{.5 0 1}{#1}}}\par\nobreak\medskip}
\def\subsect#1\par{\subsectskip\bookmark0{#1}\subsecty{#1}}
\long\def\x#1 #2\par{\hangindent2\parindent%
\mark{0 0 1}\rgboo{0 0 1}{\bf Exercise #1}\\#2%
\par\rgboo{0 0 0}\mark{0 0 0}}
\def\refs{\bigskip\noindent{\bf \rgbo{0 .5 1}{REFERENCES}}\par\nobreak\medskip
	\frenchspacing \parskip=0pt \refrm \baselineskip=1.23em plus 1pt
	\def\ital##1Õ{{\refit##1\/}}}
\long\def\twocolumn#1#2{\hbox to\hsize{\vtop{\hsize=2.9in#1}
	\hfil\vtop{\hsize=2.9in #2}}}


\twelvepoint
\font\bigger=cmbx12 \sca2
\font\biggerer=cmb10 \sca5
\font\biggest=cmssdc10 scaled 2986
 \sca5

 \sca3


\def Ü{\relax\ifmmode\Rightarrow\else\expandafter\subsect\fi}
\def Û{\relax\ifmmode\Leftrightarrow\else\expandafter\sect\fi}
\def Ú{\relax\ifmmode\Leftarrow\else\expandafter\chap\fi}

\def\itemize#1 {\item{\bf#1}}
\def\itemizze#1 {\itemitem{\bf#1}}
\def\itemutem{\par\indent\indent \hangindent3\parindent \textindent}
\def\itemizzze#1 {\itemutem{\bf#1}}
\def ª{\relax\ifmmode\leftrightarrow\else\itemizze\fi}
\def Á{\relax\ifmmode\gets\else\itemizzze\fi}

\def\tat#1#2#3#4{{\textstyle\left({#1\atop#2}{#3\atop#4}\right)}}

\def\¢{\ominus}
\def\A{{\cal A}}  \def\B{{\cal B}}  \def\C{{\cal C}}  \def\D{{\cal D}}
    \def\G{{\cal G}}

\def\Ä{\varphi}  \def\¿{\varpi}	\def\Ï{\vartheta}

\def ò{\relax\ifmmode\cdots\else\dotfill\fi}


\def\cvrule{\rgbo{0 .5 1}{\vrule}}
\def\chrule{\rgbo{0 .5 1}{\hrule}}
\def\boxit#1{\leavevmode\thinspace\hbox{\cvrule\vtop{\vbox{\chrule%
	\vskip3pt\kern1pt\hbox{\vphantom{\bf/}\thinspace\thinspace%
	{\bf#1}\thinspace\thinspace}}\kern1pt\vskip3pt\chrule}\cvrule}%
	\thinspace}
\def\Boxit#1{\noindent\vbox{\chrule\hbox{\cvrule\kern3pt\vbox{
	\advance\hsize-7pt\vskip-\parskip\kern3pt\bf#1
	\hbox{\vrule height0pt depth\dp\strutbox width0pt}
	\kern3pt}\kern3pt\cvrule}\chrule}}




\def\today{\ifcase\month\or
 January\or February\or March\or April\or May\or June\or July\or
 August\or September\or October\or November\or December\fi
 \space\number\day, \number\year}

\parindent=20pt
\newskip\normalparskip	\normalparskip=.7\medskipamount
\parskip=\normalparskip	




\catcode`\|=\active \catcode`\<=\active \catcode`\>=\active 
\def|{\relax\ifmmode\delimiter"026A30C \else$\mathchar"026A$\fi}
\def<{\relax\ifmmode\mathchar"313C \else$\mathchar"313C$\fi}
\def>{\relax\ifmmode\mathchar"313E \else$\mathchar"313E$\fi}


%
%
%
%
%
%
%

\def\thetitle#1#2#3#4#5{
 \def\titlefont{\biggest} \font\footrm=cmr10 \font\footit=cmti10
  \twelverm
	{\hbox to\hsize{#4 \hfill YITP-SB-#3}}\par
	\vskip.8in minus.1in {\center\baselineskip=2.2\normalbaselineskip
 {\titlefont #1}\par}{\center\baselineskip=\normalbaselineskip
 \vskip.5in minus.2in #2
	\vskip1.4in minus1.2in {\twelvebf ABSTRACT}\par}
 \vskip.1in\par
 \narrower\par#5\par\unnarrower\vskip3.5in minus3.3in\eject}
\def\paper\par#1\par#2\par#3\par#4\par#5\par{
	\thetitle{#1}{#2}{#3}{#4}{#5}} 
\def\author#1#2{#1 \vskip.1in {\twelveit #2}\vskip.1in}
\def\YITP{C. N. Yang Institute for Theoretical Physics\\
	State University of New York, Stony Brook, NY 11794-3840}
\def\WS{Warren Siegel\footnote{$*$}{
	\pdflink{mailto:siegel@insti.physics.sunysb.edu}\\
	\pdfklink{http://insti.physics.sunysb.edu/\~{}siegel/plan.html}
	{http://insti.physics.sunysb.edu/\noexpand~siegel/plan.html}}}


\pageno=0

\paper

{\rgb{0.5 0 0.8}{First-quantized N=4 Yang-Mills}}

\author{Machiko Hatsuda\footnote{$ÿ$}{%
	\pdflink{mailto:mhatsuda@post.kek.jp}}}
	{Theory Division, 
	High Energy Accelerator Research Organization (KEK)\\
	Tsukuba, Ibaraki, 305-0801, Japan\\
	and\\ 
	Urawa University\\
	Saitama 336-0974, Japan}
\vskip.2in
\author{Yu-tin Huang\footnote{${}^{\ddagger}$}{%
	\pdflink{mailto:yhuang@grad.physics.sunysb.edu}} 
	and \WS}\YITP

08-47,âKEK-TH-1295

December 24, 2008

We present simple first-class-constraint actions and BRST operators for the 4D N=4 superparticle.  Each generation of ghosts has 2 super-indices, like the coordinates.  The constraints that define projective superspace close in a super Yang-Mills background due to the presence of a (non-central) U(1) generator in the algebra.

\pageno=2

Û3 1. Summary

Ü1.1.  Introduction

N=4 super Yang-Mills is the simplest four-dimensional quantum field theory in terms of properties relating to symmetry, finiteness, vanishing of amplitudes, resummation, etc.  However, there is still no tractable formalism for calculating its amplitudes that directly incorporates these features.  

There has been some success with methods based on S-matrices.  Being on shell, they are necessarily focused on infrared structure; their utility for studying properties relating to confinement is not clear.  They include: (1) twistor-based recursion relations [1], which give simple expressions for tree amplitudes, but have difficulty with loops, (2) unitarity-based methods [2], and (3) string-based methods [3], which have yielded similar results for bosonic factors, but are inherently infrared due to their basis on (the leading order of) the JWKB expansion.

More general approaches incorporate the full off-shell supersymmetry manifestly, but prefer the ten-dimensional theory, showing no advantages unique to four dimensions:  (1) Pure spinors [4] have complicated loop insertions (related to picture changing) that resemble BRST operators.  (There is also the related problem of the lack of a gauge-invariant classical mechanics action, and thus of the usual $b$ and $c$ ghosts.)  (2) The use of a ghost pyramid of spinor coordinates [5] has a BRST operator (following from an infinite set of constraints) that becomes complicated in the presence of a background (although it can be simply truncated in applications so far), and its viability at higher loops is still being investigated.

In this paper we present the ingredients for a new formulation of this theory based on N=4 projective superspace [6,7].  Although a complete formalism exists for describing 4D N=2 renormalizable quantum field theories in N=2 projective superspace [8] (which, however, could stand some further elucidation), little has been done for the N=4 analog at the interacting level.  (There is an N=3 harmonic formulation [9], but no amplitudes have been calculated with it. Recently, a modified N=4 harmonic superspace has been proposed [10]; however, it failed to obtain the correct propagating degrees of freedom).  However, the simplest expression for the 4-point amplitude kinematic factor (and thus presumably the amplitude to all orders, after including the usual scalar loop factors) in normal (super) spacetime (or its conjugate momentum superspace), as opposed to supertwistor space, is in projective superspace [11].  (We will present a new derivation of this result from supertwistor space below.)  This is due to the fact that the projective superfield strength is a scalar, while the chiral superfield strength, as follows from chiral supertwistor space (which is geared for MHV amplitudes), is a tensor, whose chirality holds only at the linearized level.  This suggests that a projective formulation, at the (interacting) first- or second-quantized level, would provide the simplest derivation of this result.  Also, being in spacetime as opposed to twistor space, it would directly allow an off-shell extension.

After introducing projective superspace, in the following section we discuss the construction of constraints using suitable group generators, and proceed to solve them for the simple N=0 case. In section 3 we give a simple set of first-class superconformal (super AdS) constraints for first-quantization of the superparticle (based on an earlier description for AdS${}_5ð$S${}^5$ [12]), solve them, and give the corresponding action.  From these constraints results a simple BRST operator.  There is a supersymmetric ghost ``tower" (but not a ``pyramid") for {\it all} the coordinates.  Projective superspace is found in section 4 as a first-quantized (partial) gauge choice.  It is both a (partial) unitary gauge, in that it eliminates constraints and their corresponding ghosts, and a covariant gauge.  In section 5 the projective superspace formalism for N=4 Yang-Mills is derived by the corresponding truncation from the full superspace, which is possible only with projective superspace.  Section 6 shows how four-point amplitudes are simpler in projective superspace than chiral.  We conclude with some general remarks.

Ü1.2. Superconformal

The superspace we use was described in previous papers [13,6,7].  Most of its simplifications follow from the fact that its coordinates are conveniently arranged in a single matrix.   It can be understood as a graded matrix generalization of quaternionic projective space.  As a result its superconformal transformations can be represented as fractional linear transformations:
$$ w' = (wc + d)^{-1}(wa + b) $$
or the equivalent
$$ w' = (÷aw + ÷b)(÷cw + ÷d)^{-1} $$
in terms of the (P)SU(N|2,2) group element
$$ \pmatrix{ a & c \cr b & d \cr} = {\pmatrix{ ÷d & -÷c \cr -÷b & ÷a \cr}}^{-1} $$
or in linearized form as
$$ ¶w = Œ + ºw + w© + w·w $$
where $w$ is an $(n|2)$ by $(N-n|2)$ matrix for the case of $N$ supersymmetries, where ``$n$" indicates their ``twisting":  $n=0$ (or $N$) describes (anti)chiral superspace, while $n=N/2$ describes the preferred superspace, which allows a type of reality condition because this matrix is then square, satisfying a form of hermiticity.   It then has an equal number of $Ï$'s and $ÐÏ$'s, as well as the usual $x$'s and some R-symmetry coordinates $y$:
$$ w_{A'}{}^A = \;\bordermatrix{ & a & Π\cr 
	a' & y_{a'}{}^a & Ï_{a'}{}^Œ \cr
	\vrule height 1.4em width 0em
	ÀŒ & ÐÏ_{ÀŒ}{}^a & x_{ÀŒ}{}^Œ \cr} $$

The (pseudo)unitarity of the group element implies a charge conjugation transformation [7]:  $\C w$ transforms in the same way under the group as $w$ if we define
$$ (\C w)ÿ ­ \bordermatrix{ & a' & Œ \cr
	a & -y^{-1}{}_a{}^{a'} & iy^{-1}{}_a{}^{b'}Ï_{b'}{}^º C_{ºŒ} \cr
	\vrule height 1.4em width 0em
	ÀŒ & -iC^{ÀŒÀº}ÐÏ_{Àº}{}^b y^{-1}{}_b{}^{a'} 
	& C^{ÀŒÀº}(x_{Àº}{}^º-ÐÏ_{Àº}{}^b y^{-1}{}_b{}^{b'}Ï_{b'}{}^º)C_{ºŒ} \cr} $$ 
The superconformally invariant (4D extension of the Hilbert space) inner product is then
$$ ÒA|BÔ ­ Çdw¼(\C A)(w) B(w) = ÒB|AÔ* $$
for any $A$ and $B$ that transform as half-densities
$$ dw'[A'(w')]^2 = dw[A(w)]^2 $$
where $\C A$, which transforms in the same way as $A$, is defined by the relation of complex conjugation to charge conjugation in the above inner product,
$$ (\C A)(w) ­ [det(y)]^{-str(I)} [A(\C w)]ÿ,âstr(I) = üN-2 $$

This superspace has various descriptions in terms of cosets [6] or related projections [13,7].  A manifestly superconformal description is most natural in a projective lightcone formalism [13,7]: In that approach, one would start with the coset of the superconformal group with respect to a classical (isotropy) group, and take a contraction of the latter (``projective lightcone limit"), which makes some of the original coordinates (including one from spacetime) nondynamical.  Unfortunately, the interpretation of the resulting action remains unclear.

Ü1.3. Super anti-de Sitter

In this paper we use a new approach:  We first formulate the superparticle theory in the group space of the isometry group.  We then choose first-quantized gauge conditions, which generate the isotropy group.  Thus the ``full" superspace of the isometry group is reduced to its projective coset.  This approach has the advantage that before gauge fixing the Yang-Mills background can be constructed with covariant derivatives, which require the full superspace, while after gauge fixing the theory can be quantized using just projective superspace, which is all the Yang-Mills prepotential requires.  Also, the fact that the action for the N=2 vector multiplet in harmonic superspace is nonlocal in the internal coordinates is analogous to Coulomb-like interactions, suggesting such spaces are the result of partial gauge fixing from larger superspaces with local actions.  In general, reduction of the number of fermionic coordinates is useful for quantization because only one quarter of those of the full superspace are physical; any unphysical coordinates must be canceled by ghosts.  However, in such spaces nonrenormalization theorems are not obvious.

As the isometry group we choose the super-anti de Sitter group (in four dimensions).  The isotropy group is the super-anti de Sitter group in one lower dimension (three), up to questions of signature.  (Some Wick rotation is involved, since we really want 3D de Sitter symmetry, not 3D anti-de Sitter, to get anti-de Sitter space SO(3,2)/SO(3,1), but only the anti-de Sitter symmetry can be supersymmetrized.  This Wick rotation leads to modified reality conditions, which always occur in projective space [8,7].)  Although the manifest symmetry is only super-anti de Sitter, the superconformal invariance of super Yang-Mills in D=4 guarantees the result is directly applicable to flat space.  (Pure spinors have also been used to describe 4D N=4 Yang-Mills in super AdS, but using the maximal bosonic isotropy group [14])

The relevant cosets are then
$$ {\rm OSp(N|4) \over OSp(n|2)OSp(N-n|2)} $$

\noindent which can readily be seen to yield the rectangle of coordinates given above. (This isotropy group was also found in the projective lightcone approach. The case n=0 of this coset, namely ${\rm OSp(N|4)\over OSp(N|2)Sp(2)}$, was used in [15] to describe self-dual supergravity).

We can also take a contraction of the isometry, a graded generalization of the contraction used to obtain the Poincar«e group from the anti-de Sitter group:  In terms of the algebra
$$ [H,HÕ = H,â[H,G/HÕ = G/H,â[G/H,G/HÕ = H $$
we simply drop the right-hand side of the last equation.  The result is
$$ {\rm I[OSp(n|2)OSp(N-n|2)]\over OSp(n|2)OSp(N-n|2)} $$
where the ``I(nhomogeneous)" part is just translations with respect to the coordinates of the rectangle described above.  The coset is now Abelian, and consists of the usual ``supertranslations" of the projective group: spacetime translations, half of the supersymmetries, and part of the R-symmetry.  However, although the part of the isometry group acting on Lorentz (spinor) indices is just the Poincar«e group, the full group is not the super Poincar«e group, because the isotropy group is unchanged:  The last consists of the Lorentz group, a subgroup of R-symmetry, and the sum of the other half of the supersymmetries and the corresponding S-supersymmetries.  So we have the usual flat spacetime, but not the usual flat full superspace.  Since the coset space is our projective superspace on which our superfields depend, while the isotropy group is the tangent space, this means we have a flat (and torsion-free) coordinate space with a curved tangent space, the opposite of the usual.

We will use both these cosets below (more or less simultaneously, since it's easy to see how to contract the former to the latter).  Both isometry groups are subgroups of the superconformal group.

Û2 2. Groups without cosets

Ü2.1. Constraints

All the constraints we consider in the following sections are first-class.  All the actions are of the form
$$ S = Çd ¼L,ââL = -Àz^î p_î + Â^i \G_i $$
for superspace coordinates $z$ and their conjugate momenta $p$, and constraints $\G$ and their Lagrange multipliers $Â$, all as functions of the worldline parameter $ $.

For reasons mentioned above, we define our ``full" superspace as the entire supergroup space, rather than just a coset.  The free field equations of the theory are expressed at the first-quantized level as constraints quadratic in the group generators.  (This is ``dual" to writing them in the same form in terms of covariant derivatives.)  We begin by reducing the (symmetrized) square of the generators:  The three general (finite-dimensional) cases are for superconformal in D=3,4,6 (or 1 more dimension for super-AdS),
 \def\inny#1{õ \hbox to 0pt{\kern-.6em\raise-.4ex\hbox{$^{#1}$}}}
 \def\dumbõ#1{\mkern-3.5mu \inny{#1}}
 \def\dumberõ#1{\llap{\raise-1.5ex\hbox{$\inny{#1}$}}}
 {\hbadness=10000
$$ \openup1\jot \li{{\rm OSp(N|4):} \left( \upõ2{\õ2} ° \upõ2{\õ2} \right)_S & = 
	\upõ4{\õ4} ¢ \upõ2{\õ2\õ2} ¢ \upõ1{\õ1\õ1} ¢ \bullet \cr
{\rm (P)SU(N|2,2):} \left( \upõ1{\õ1\dumbõ{\bullet}} ° \upõ1{\õ1\dumbõ{\bullet}} \right)_S& = 
	\upõ2{\õ2\dumbõ{\bullet}\dumberõ{\bullet}} ¢ 
	\upõ2{\dumbõ{}\dumberõ{\bullet}\dumbõ{}\dumberõ{\bullet}} ¢
	\upõ1{\õ1\dumbõ{\bullet}} ¢ \bullet \cr
{\rm OSp*(8|2N):} \left( \upõ2{\õ2} ° \upõ2{\õ2} \right)_S & = 
	\upõ2{\õ2\õ2} ¢ \upõ4{\õ4} ¢ \upõ1{\õ1\õ1} ¢ \bullet \cr} $$
}(OSp has a real defining representation, OSp* has pseudoreal; thus the former has bosonic subgroup SO(N)Sp(4), while the latter has SO*(8)USp(2N).)  In each case we have listed the 4 irreducible representations in the following order:

\narrower\parskip=0pt\noindent
\item{(1)} nonvanishing on shell, most symmetric in spacetime spinor indices;
\item{(2)} vanishing for superconformal only, most antisymmetric in spacetime spin\-or indices, includes flat Klein-Gordon;
\item{(3)} vanishing for both superconformal and AdS, single supertrace, includes Pauli-Lubanski;
\item{(4)} vanishing for both superconformal and AdS, double supertrace (Casimir, with a dot for its singlet tableau), AdS Klein-Gordon.

\unnarrower\parskip=\normalparskip

We use graded symmetrization, so ``symmetric" in the tableaux means symmetric in the former label of the group, since in the first and last cases that has the symmetric metric.  For the cases of interest in D=4, this means that R-indices are considered as bosonic, and Lorentz spinor indices as fermionic, for purposes of sign factors from reordering indices.  So, e.g., $str(M_\A{}^\B)=+M^\A{}_\A$.  For the unitary case, dots in boxes refer to the complex conjugate representation; ordering of the dotted block with respect to the undotted block is arbitrary.  In the cases where the ranges of the bosonic and fermionic indices are the same (N=4), supertracelessness is undefined ($str(I)=0$), so the 3rd constraint implies the 4th, and the 3rd is implied by the 1st (for the latter 2 cases) or the 2nd (for the former 2 cases).  By ``vanishing", we mean up to constants, for the case of vanishing superhelicity.  (Nonvanishing superhelicity can be described by introducing ``spin" operators, in addition to these ``orbital" ones defined in terms of just the supergroup coordinates.)

To see that this classification of (quadratic) constraints is consistent with the usual identification of the superconformal mass shell, we evaluate them in the supertwistor representation, which exists for D=3,4,6:  The generators $öG$ in terms of the supertwistors $½$ are
$$ \li{ D=3 & : öG_{\A\B} = ü[½_\A,½_\BÕ,ââÓ½_\A,½_\B] = ú_{\A\B} \cr
D=4 & : öG^\A{}_\B = ü[½^\A,н_\BÕ,ââÓн_\A,½^\B] = ¶_\A{}^\B,ââââh = ü[½^\A,н_\AÕ \cr
D=6 & : öG_{\A\B} = üÓ½^a{}_\A,½_{a\B}],â¼[½_{a\A},½_{b\B}Õ = C_{ab}ú_{\A\B},ââ
	h_{ab} = üÓ½_a{}^\A,½_{b\A}] \cr} $$
with indices $\A,\B$ in the defining representation (and defining SU(2) indices $a,b$ for D=6), where $h$ is the superhelicity (generating the little group U(1) for D=4, or SU(2) for chiral D=6), which is set to vanish in our case.  (For D=4 we have given the U(N|2,2) generators; in coordinate representations, only (P)SU(N|2,2) need be defined.  Note that twistors are essentially $©$-matrices for OSp, or creation/annihilation operators for U, satisfying graded ÓantiÕcommutation relations; thus the bosonic ones anticommute with the fermionic ones.)  

Substitution of these representations into the corresponding constraints numbers 2,3,4 above shows they vanish up to constants for vanishing superhelicity, and do not vanish for number 1. 

Once the correct constraints have been identified, it's more convenient (for purposes of applying isotropy conditions or coupling background fields) to express the constraints in their dual form in terms of (free) covariant derivatives ($öG£d$). 

Ü2.2. N=0

For the case N=0, we can examine arbitrary dimensions, with the generators carrying vector indices; then we have
$$ {\rm SO(D,2):} \left( \upõ2{\õ2} ° \upõ2{\õ2} \right)_S = 
	\upõ2{\õ2\õ2} ¢ \upõ1{\õ1\õ1} ¢ \upõ4{\õ4} ¢ \bullet $$
(The ordering is as above, but the roles of symmetry and traces have changed.)

We now outline the solution of the constraints.  In the conformal case, the constraints are, in terms of SO(D$-$1,1) indices,

$$ \def\normalbaselines{\advance\baselineskip1\jot}
	\matrix{ dim & \upõ1{\õ1\õ1} & \upõ4{\õ4} & \bullet \cr
	2 & P^2 & & \cr
	1 & S_m{}^n P_n + w P_m & S_{[mn}P_{p]} & \cr
	0 & S_m{}^p S_{pn} + K_{(m}P_{n)} - tr  & wS_{mn} + K_{[m}P_{n]} & \cr
	0 & S^2 + (D+1)w^2 + (D-2)KÉP & S_{[mn}S_{pq]} & üS^2 - w^2 - 2KÉP \cr
	-1 & S_m{}^n K_n - wK_m & S_{[mn}K_{p]} & \cr
	-2 & K^2 & & \cr} $$
for momentum $P$, spin $S$, scale weight $w$, and covariant derivative for conformal boosts $K$.  The dimension-2 constraint allows us to choose a lightcone frame to make the analysis simpler.  The dimension-1 constraints then imply $S=w=0$ (assuming the momentum is not identically 0).  The surviving dimension-0 constraints then imply $K=0$.  So we are left with a massless scalar.

In the AdS case, in terms of SO(D,1) indices, we have
\par\nobreak
$$ \def\normalbaselines{\advance\baselineskip1\jot}
	\matrix{ \upõ4{\õ4} & \bullet \cr
	& P^2 - üS^2\cr
	S_{[mn}P_{p]} & \cr
	S_{[mn}S_{pq]} & \cr} $$

\noindent We again find $S=0$, describing a scalar.

\sectskip\bookmark3{3. AdS4 supergroup space}\cross=1
	\xdef\righthead{\rgbo{1 0 1}{3. AdS${}_4$ Supergroup space}}\vbox{\bigger\offinterlineskip
	\line{\cdotfill¼\hphantom{{3. AdS${}_{\textstyle\bf 4}$ supergroup space}}\cdotfill}
	\line{\cdotfill¼{\rgbo{1 0 1}{3. AdS${}_{\textstyle\bf 4}$ supergroup space}}\cdotfill}}\par\nobreak\medskip	

Ü3.1. Constraints

We now examine the constraints in more detail for the cases of interest for D=4: (P)SU(N|2,2) for superconformal, and OSp(N|4) for super-AdS.  Conveniently, D=4 is the only case for which there are classical supergroups for all N that can be applied for both superconformal and super-AdS.  As a result, a supertwistor analysis can be applied for super-AdS as well as for superconformal (using the same supertwistors).

For the superconformal group we find from the analysis of the previous section, substituting the supertwistor expression,
 {\hbadness=10000
$$ superconformalâ\upõ2{\dumbõ{}\dumberõ{\bullet}\dumbõ{}\dumberõ{\bullet}} ¢
	\upõ1{\õ1\dumbõ{\bullet}} ¢ \bullet:ââ
	ö\G_{\A\B}{}^{\C\D} ­
	öG_{(\A}{}^{(\C}öG_{\B]}{}^{\D]} - ü¶_{(\A}{}^\C ¶_{\B]}{}^\D = 0 $$
}The Kronecker $¶$ term can be considered as arising from ``normal ordering".  

However, the analog in the AdS case does not take a similar form, although the constraints there are a subset of (a linear combination of) the superconformal ones, because the OSp(N|4) generators themselves result from (graded) antisymmetrization of the (P)SU(N|2,2) ones:
$$ öG_\A{}^\Bâ£âG_{\A\B} ­ öG_{[\A|}{}^\C ú_{\C|\B)} $$
where $ú$ is the graded-symmetric OSp(N|4) metric.  Consequently the maximal subset of the above constraints that can be expressed in terms of OSp(N|4) generators requires a supertrace on the indices of the above constraints, in agreement with our previous analysis:
$$ super-AdSâ\upõ1{\õ1\õ1} ¢ \bullet:ââ
	\G_{\A\B} ­ G_{(\A|}{}^\C G_{\C|\B]} + str(I)ú_{\A\B} = 0 $$
The superhelicity is again required to vanish.  (Note the ``anomalous" $ú$ term vanishes for N=4.)  Similar constraints were proposed previously for the (10D) superparticle on AdS${}_5ð$S${}^5$ [11], using the superconformal isometry group.  

Ü3.2. Lightcone gauge

The constraints are most easily solved in a lightcone-type decomposition.  First, it's useful to identify how the constraints relate to the more reducible superconformal constraints.  It will only be necessary to look at those constraints that do not include terms with covariant derivatives corresponding to conformal boosts and S-supersymmetry, since those constraints can be applied to arbitrary massless representations of supersymmetry [16], of which the minimal representation appearing in first-quantization is a special case.  (So the rest are redundant, at least after choosing conformal boosts and S-supersymmetry as the isotropy group.)  Separating out the PSU(N|2,2) indices as $\A=({\rm a},Œ,ÀŒ)$ (where a = $(a,a')$), those constraints are

$$ \def\normalbaselines{\advance\baselineskip1\jot}
	\matrix{ \hfill\A & ­ & \f14ö\G_{Œº}{}^{À©À¶}C^{ºŒ}ÐC_{À¶À©} & = & p^{ŒÀŒ}p_{ŒÀŒ} \hfill\cr
	\hfill\B_{{\rm a}Œ} & ­ & üö\G_{{\rm a}Œ}{}^{À©À¶}ÐC_{À¶À©} 
		& = & p_Œ{}^{ÀŒ}й_{{\rm a}ÀŒ} \hfill\cr
	\hfillÐ\B^{{\rm a}ÀŒ} & ­ & üö\G_{Œº}{}^{{\rm a}ÀŒ}C^{ºŒ} 
		& = & p^{ŒÀŒ}¹^{\rm a}{}_Œ \hfill\cr
	\hfill\C_Œ{}^{ÀŒ} & ­ & ö\G_Œ{}^\B{}_\B{}^{ÀŒ} 
		& = & S_Œ{}^º p_º{}^{ÀŒ} - ÐS^{ÀŒÀº}p_{ŒÀº} -¹^{\rm a}{}_Œ й_{\rm a}{}^{ÀŒ} \hfill\cr
	\hfill\C_{int,{\rm a}Œ}{}^{{\rm b}ÀŒ} & ­ & ö\G_{{\rm a}Œ}{}^{{\rm b}ÀŒ} 
		& = & t_{\rm a}{}^{\rm b}p^{ŒÀŒ} +¹^{\rm b}{}_Œ й_{\rm a}{}^{ÀŒ} \hfill\cr
	\hfill\C_{}{}^{\rm ab}{}_{Œº} & ­ & ö\G_{Œº}{}^{\rm ab}
		& = & ¹^{\rm a}{}_Œ ¹^{\rm b}{}_º \hfill\cr
	\hfill\C_{Ѝ,\rm ab}{}^{ÀŒÀº} & ­ & ö\G_{\rm ab}{}^{ÀŒÀº}
		& = & й_{\rm a}{}^{ÀŒ} й_{\rm b}{}^{Àº} \hfill\cr
	\hfill\D_Œ{}^{ÀŒ} & ­ & ü(ö\G_{Œº}{}^{ºÀŒ}+ö\G_{ŒÀº}{}^{ÀŒÀº}) 
		& = & S_Œ{}^º p_º{}^{ÀŒ} +ÐS^{ÀŒÀº}p_{ŒÀº} +wp_Œ{}^{ÀŒ} \hfill\cr} $$
	
\noindent where we have labeled the constraints as in [16].  Note that in D=4 the Pauli-Lubanski equation $\C$ is equivalent to the $\D$ constraint in the presence of the $\B$ constraint and the Klein-Gordon equation $\A$.  

We now compare these to the OSp(N|4) constraints:  In terms of
$$ d^{\A\B} = \bordermatrix{ & {\rm b} & º & Àº \cr
	{\rm a} & t^{\rm ab} & ¹^{{\rm a}º} & й^{{\rm a}Àº} \cr
	Œ & -¹^{{\rm b}Œ} & S^{Œº} & p^{ŒÀº} \cr
	ÀŒ & -й^{{\rm b}ÀŒ} & p^{ºÀŒ} & ÐS^{ÀŒÀº} \cr} $$
we have
$$ 0 = \G^{\A\B} ­ d^{\A\C}d^{\B\D}ú_{\D\C} = $$
{\hfuzz=8pt
$$ \hskip-4pt\pmatrix{ 
	t^{\rm ac}t^{\rm b}{}_{\rm c} +¹^{{\rm a}©}¹^{\rm b}{}_© 
			+й^{{\rm a}À©}й^{\rm b}{}_{À©}
		& t^{\rm ac}¹^º{}_{\rm c} -¹^{{\rm a}©}S^º{}_© -й^{{\rm a}À©}p^º{}_{À©}
		& t^{\rm ac}й^{Àº}{}_{\rm c} -¹^{{\rm a}©}p^{Àº}{}_© -й^{{\rm a}À©}ÐS^{Àº}{}_{À©} \cr
	... & ¹^{Œ\rm c}¹^º{}_{\rm c} -S^{Œ©}S^º{}_© -p^{ŒÀ©}p^º{}_{À©}
		& ¹^{Œ\rm c}й^{Àº}{}_{\rm c} -S^{Œ©}p^{Àº}{}_© -p^{ŒÀ©}ÐS^{Àº}{}_{À©} \cr
	... & ... & й^{ÀŒ\rm c}й^{Àº}{}_{\rm c} -p^{ÀŒ©}p^{Àº}{}_© -ÐS^{ÀŒÀ©}ÐS^{Àº}{}_{À©} \cr} $$
}
$$ ® \pmatrix{ ? & \B^{{\rm a}º} & Ð\B^{{\rm a}Àº} \cr
	... & C^{Œº}\A & \C^{ŒÀº} \cr
	... & ... & ÐC^{ÀŒÀº}\A \cr} $$
where ``$...$" is proportional to the transposed element, ``$®$" refers to extra terms, and we have ignored symmetrization of indices, which produces terms linear in generators.  Note that extra signs from reordering of super-indices are implicit:  For example, in the supertrace of indices in the definition of the OSp constraints, there is a factor of $(-1)^{\B\D}$, which is $-1$ if both indices are Sp(4) and $+1$ otherwise, because supertraced indices belong next to each other.  (We could also just use the graded symmetry of the second $d$ factor, but we want to use notation that applies to the general case, where $d$ has no symmetry.)  The constraint ``Ê?Ê" will be found to be the square of the lightcone part of the $\C_{int}$ constraint.

To analyze these constraints it's instructive to look first at the case N=0:  In the superconformal case, we can solve the $\A$ constraint as usual.  The $\C$ constraint is then the usual Pauli-Lubanski equation for vanishing helicity:  We can thus set the spin operators $S^{Œº}$ and $ÐS^{ÀŒÀº}$ to vanish.  (The components of the spin not explicitly set to vanish by this equation do not appear, and so can be eliminated from the theory by unitary transformation, or equivalently by a gauge condition for the gauge transformation generated by this equation.)  The $\D$ constraint does the same if we set the conformal weight $w=0$:  It's the same as the Pauli-Lubanski equation except for a (Hodge) duality transformation on the spin (and in general also switching helicity with conformal weight).  The AdS case is similar except for the extra terms in $\A$; but these drop out after solving the Pauli-Lubanski equation.

We then choose a lightcone Lorentz frame
$$ p^{ŒÀŒ} = \bordermatrix{ & \rdt + & \rdt - \cr + & p^+ & 0 \cr - & 0 & p^- \cr} $$
For general N, the $\A$ and $\B$ constraints are used to solve for $p^{-À-}$, and $¹_{\rm a}^-$ and $й_{\rm a}^{À-}$, as usual.  The $\C$ constraint then determines $S^{Œº}$ and $ÐS^{ÀŒÀº}$.  (Again a $\D$ constraint is unnecessary.)  Now the ? constraint will perform a similar function for $t_{\rm ab}$:  After plugging in the solution for $¹_{\rm a}^-$ and $й_{\rm a}^{À-}$, it becomes
$$ ÷t^{\rm ac}÷t^{Ê\rm b}{}_{\rm c} = 0,ââ÷t^{\rm ab} = t^{\rm ab} 
	- {1\over p^+}й^{\rm a\rdt +}¹^{\rm b+} $$
We recognize $÷t^{\rm ab}$ as proportional to the superconformal $\C_{int}^{{\rm ab}+À+}$.  Since the internal space is compact, the vanishing of the square of this operator implies its own vanishing.  (In particular, we see all the Casimirs of these modified group generators vanish.)  Thus the AdS constraints are equivalent to the larger superconformal set:  They yield the supertwistor representation.

Ü3.3. BRST

Isotropy constraints (really gauge conditions) are expressed in terms of covariant derivatives (since they preserve the global symmetry), so from now on we also represent the quadratic constraints (field equations) in terms of them, also.  (The supertwistor representation of the previous subsection applies only to the group generators.)  The covariant derivatives are a subset of those for the superconformal group.  The explicit form of the latter has been given previously; we won't need them here (only their algebra).  

In matrix notation, these constraints are
$$ dúd^T = d^T úd = 0 $$
(for graded transpose ``$Ê{}^TÊ$").  The most interesting things about these constraints are that:  (1) Their index structure is that of a matrix, as for the covariant derivatives $d$ themselves, except perhaps for the symmetry on their two indices; and (2) the same is true for their reducibility condition, and the reducibility of the reducibility condition, etc.  The net result is that the complete minimal BRST operator can be written in the simple form (matrix multiplication with metric, and trace, implied)
$$ Q = Ý_{m,n=0}^\infty c_{m+n+1}b_m b_n + f... $$
where the indices label the ghost generation,
$$ b_0 = d $$
and ``$f...$" denotes structure-constant terms.  (We won't need those for the contracted projective case.)


In the present case $d$ is graded antisymmetric on its 2 indices.  We then have 
$$ d^T = -d $$
$$ (dúd)^T = +(dúd) $$
for the constraints $\G_1­dúd=0$.  Then the reducibility conditions are
$$ \G_2 ­ dú\G_1 - \G_1 úd = +\G_2^T = 0 $$
$$ \G_3 ­ dú\G_2 + \G_2 úd = -\G_3^T = 0 $$
etc., where the sign for the symmetry of $\G_n$ alternates as $-++--++--...$ .  Explicitly, with $\G_0­d$,
$$ \G_{n+1} ­ dú\G_n +(-1)^n\G_n úd = (-1)^{n(n-1)/2}\G_{n+1}^T = 0 $$
Using this construction for the BRST operator, and including terms for closure ($Q^2=0$) leads to the above expression for the BRST operator, where
$$ c_n = -(-1)^{n(n+1)/2}c_n^T $$
and similarly for $b$.

No nonminimal degrees of freedom are needed; we can choose a ``temporal" gauge for the first-quantized gauge fields, as is standard for D=1 and 2 (because it doesn't break worldsheet or spacetime Lorentz invariance).  Thus, there is only a (``1D") tower of ghosts [13] (for all of $x$, $Ï$ and $y$), as opposed to the (``2D") pyramid of ghosts (for just $Ï$) for the approach of that name.

Û3 4. N=4 projective superspace

Ü4.1. Projective gauge

In the previous section we analyzed the first-quantized theory on shell by simultaneously solving all the constraints explicitly and choosing a lightcone gauge for the symmetry generated by the constraints.  We can instead solve a subset of the constraints and choose their corresponding gauges in such a way as to manifestly preserve Lorentz covariance.  This can be achieved in a way that is equivalent to completely eliminating some of the coordinates (a subset of those eliminated in the lightcone gauge).  Since the algebra of gauge conditions must close, this is the same as choosing an isotropy subgroup.
Then the isotropy group can be used to reduce the original constraints, eliminating constraints, or terms in constraints, that vanish off shell as a consequence of the vanishing of the isotropy covariant derivatives themselves.  This leads to the cosets of subsection 1.3, corresponding to projective superspace.

To treat these cosets, we divide the range of OSp(N|4) indices in half as
$$ \A = (A,A') $$
for the two OSp($ü$N|2)'s.  The constraints resulting from dropping isotropy covariant derivatives $d_u$, leaving just the projective ones $d_w$, will have a similar form as before, but the indices will be reduced from OSp(N|4) to (one of the) OSp($ü$N|2):  $d_w$ has the index structure $d_A{}^{A'}$.

Indices are contracted with one of the OSp($ü$N|2) metrics (or its inverse),
$$ ú^{AB} = (¶^{ab},C^{Œº}) ,âú_{A'B'} = (¶_{a'b'},ÐC_{ÀŒÀº}) ;ââ
	C_{Œº} = ÐC_{ÀŒÀº} = -C^{Œº} = -ÐC^{ÀŒÀº} = \tat0i{-i}0 $$
(If we want to keep track of dimensional analysis, we can include a factor of the anti-de Sitter radius $1/R$ with the Kronecker $¶$'s, but being careful to distinguish the inverse metrics, where $R$ appears instead of $1/R$.)
  
In the explicit form for the BRST operator (which takes the same form as above, but with different symmetry for the matrices, as discussed below), the algebra for $d_w$ closes only on $d_u$, so $Q^2=0$ modulo such terms.  A separate term to enforce $d_u=0$ can easily be added, along with the corresponding terms for closure of the $d_w d_w$ algebra on $d_u$.  (Similar remarks apply to adding a Lagrange multiplier term for $d_u$ to the Hamiltonian.)  Alternatively, we can work just in the contracted coset space, and $d_u$ can be ignored altogether.  If we use the contracted coset, the $d_w$ are simply partial derivatives with respect to $w$ (up to factors of isotropy coordinates $u$, which can be ignored upon restriction to the coset space).

Ü4.2. Lightcone gauge again

First we write out the different Lorentz pieces of the constraints:  In terms of
$$ d_A{}^{A'} = \bordermatrix{ & a' & ÀŒ \cr
	a & t_a{}^{a'} & й_a{}^{ÀŒ} \cr
	Œ & ¹_Œ{}^{a'} & p_Œ{}^{ÀŒ} \cr} $$
we have
$$ \G_{AB} ­ d_A{}^{A'}d_B{}^{B'}ú_{B'A'} = \bordermatrix{ & b & º \cr
	a & t_a{}^{a'}t_{ba'} + й_a{}^{ÀŒ}й_{bÀŒ} & t_a{}^{a'}¹_{ºa'} - й_a{}^{ÀŒ}p_{ºÀŒ} \cr
	Œ & ... & ¹_Œ{}^{a'}¹_{ºa'} - p_Œ{}^{ÀŒ}p_{ºÀŒ} \cr} = 0$$
(with the usual signs for a fermionic index interrupting the contraction of two fermi\-onic indices) and similarly for the complex-conjugate constraints $Ð\G^{A'B'}$.  In particular we have, from $\G_{Œº}$, $\G^a{}_a$, and their complex conjugates,
$$ p^2 = ¹^2 = й^2 = -t^2 $$
(plus one redundant equation).  

From $\G^{+a}$ and its complex conjugate we find
$$ ¹^{-a'} = {1\over p^+}it^{aa'}й_a{}^{\rdt +} $$
and its complex conjugate.  ($\G^{-a}$ and its complex conjugate are redundant.)  This tells us
$$ ¹^2 = {1\over p^+}t^{aa'}й_a{}^{\rdt +}¹^+{}_{a'} $$
We then have, from the remaining constraint $\G_{ab}$ and complex conjugate,
$$ öt_a{}^{a'}öt_{ba'} = öt^{a}{}_{a'}öt_{ab'} = 0 $$
defining
$$ öt_{aa'} ­ t_{aa'} - {1\over p^+}й_a{}^{\rdt +}¹^+{}_{a'} $$
This expression is the independent piece of the constraint from the bigger superconformal set,
$$ p_{ŒÀŒ}t_{aa'} - й_{aÀŒ}¹_{Œa'} = 0 $$

Since the vanishing of the square of a Hermitian operator implies the vanishing of the operator, we find 
$$ öt_{aa'} = 0 $$
(This is clear on the original coset, since the internal space is compact, so there is no ambiguity in normalization of states.  However, things might be more subtle on the contracted coset.)  The hermiticity of this operator follows from the fact that it is a piece of the superconformal field equations, which can be expressed in terms of group generators (instead of covariant derivatives), which are by definition Hermitian.

It then follows that
$$ p^2 = ¹^2 = й^2 = -t^2 = 0 $$
($t^2=0$ does not imply $t_{aa'}=0$, since $t$ is not Hermitian with respect to the charge conjugation that defines the inner product.)  So $x$ dependence is determined by the Klein-Gordon equation (as usual in the lightcone formalism), $y$ dependence is completely determined, and $Ï$ dependence is determined in terms of half the original $Ï$'s, i.e., 1/4 of those of the full superspace, as usual.

Ü4.3. Counting

The results of subsection 3.3 for counting ghosts can be applied directly to the 4D case by using OSp(4|N) indices, dividing their ranges in half, and dropping irrelevant blocks.  The result is as above for odd $n$, except both indices are primed or both unprimed, while for even $n$ we have mixed indices:
$$ \cases{ c_{2n+1,AB},¼c_{2n+1,A'B'} & where $c_{2n+1} = (-1)^n c_{2n+1}^T$ \cr
	c_{2n,AB'} & \cr} $$
Thus the symmetry has a cycle of 4, going as asymmetric, (twice) graded symmetric, asymmetric, (twice) graded antisymmetric.

We can now count the naive  effective number of modes for any of $x$, $Ï$, $y$.  Infinite sums can be defined, e.g., by regularization:
$$ (1+z)^{-1} = 1 - z +z^2 - ...âÜâ1 - 1 + 1 - ... = ü  $$
We then have for each variable (remember $w$ carries OSp($ü$N|2) indices)
$$ \li{ 
x:â& 4 - 2É1 + 4 - 2É3 + ... \cr
	& = (1 - 1 + 1 - ...)É4 + (1 - 1 + 1 - ...)É2 \cr
	& = 3 = D-1 \cr
Ï:â& 2N - 2N + 2N - 2N + ... \cr
	& = (1 - 1 + 1 - ...)É2N \cr
	& = N \cr
y:â& (üN)^2 - 2É{üN(üN+1)\over 2} + (üN)^2 - 2É{üN(üN-1)\over 2} + ... \cr
	& = (1 - 1 + 1 - ...)É\f14N^2 - (1 - 1 + 1 - ...)ÉüN \cr
	& = \f18N(N-2) \cr
	} $$
where for $x$ and $y$ we have separated the sum into averages over symmetry/antisym\-metry plus the deviations due to either.  The $x$'s (and $p$'s) have just the ``transverse" degrees of freedom $D-1$, which in the equivalent ghost-free lightcone analysis arise from the gauge choice $x^+= $ (and $p^2=0$ eliminating $p^-$).  This agrees with the usual scalar particle, which has just $x$ and 1 $c$; but here there are 2 (identical) constraints for $N=0$, resulting in reducibility to cancel 1.  For $Ï$ we also find the number of physical degrees, which is just 1/4 that of the full superspace.  However, though the $y$'s have no physical degrees of freedom, they do not cancel by this counting because they are eliminated by quadratic constraints, not linear.  (But note that net bosons and fermions cancel for $N=4$, as they do at each ghost level.  Also, because of the grading the $x$ counting is just the $N=-4$ case of the $y$ counting.)  Interestingly, for the case of OSp(n|2)OSp(N$-$n|2) with $n±üN$ the sum diverges for $y$, even with the above regularization, giving an extra term $-(n-üN)^2É(1+1+...)$.


Û3 5. N=4 super Yang-Mills

Ü5.1. Covariant derivatives

We now consider the formulation of N=4 super Yang-Mills in this projective superspace.  Since in any linearized quantum gauge theory in a background of the same gauge theory, linearized gauge invariance of the quantum theory requires the background to be on shell [15], we will here restrict ourselves to an on-shell background.  However, this background is on shell with respect to the full nonlinear field equations, which is sufficient to construct the Feynman rules:  For example, tree graphs can be derived from perturbative solutions to the classical equations of motion.  Thus, the existence of this construction, combined with the off-shell formulation of the linearized theory, should be sufficient to prove the existence of the nonlinear off-shell theory, which will be left to another paper.

To relate to the known on-shell theory in the ``full" superspace, we first consider its manifestly OSp(N|4) covariant form.  Then we will show that, after a simple redefinition, the covariant derivatives in the isotropy subspace have vanishing field strengths, allowing the coset to be taken.  

In addition to the usual 4 spacetime and 4N anticommuting coordinates, this full superspace contains also internal (bosonic) coordinates for not only the AdS R-symmetry group SO(4) but also the Lorentz group.  Of course, as for (N=0) gravity in curved space, we treat the spacetime derivatives and Lorentz (spin) generators as separate, even though the Killing vectors of AdS that generate SO(3,2) do not distinguish ``translations" from ``Lorentz transformations".  This is fully consistent with the distinction between symmetry generators and covariant derivatives, and thus the usual coset construction for Sp(4)/Sp(2)$^2$ (``left" and ``right" action on the group elements).  What is unusual here is that we introduce coordinates for the Lorentz spin, as well as corresponding components for the Yang-Mills gauge fields.  (The Yang-Mills gauge group is still the same; it is only that it is defined over a bigger manifold.)  This is already done for R-symmetry, in projective and harmonic superspace.  The Yang-Mills field strengths in these directions vanish, and thus gauges can be chosen where their gauge fields do also.  However, in some cases it may prove convenient to choose gauges where they do not, as in the usual N=2 harmonic construction [16].  An interesting example is the case of selfdual Yang-Mills, even for N=0, which is known to be analogous to N=2 projective and harmonic superspaces [15,8].  In the lightcone gauge for this theory, we separate the $+$ and $-$ components of the undotted spinor index (but not the dotted one) to solve some of the selfduality conditions as
$$ \{  á_{+ÀŒ} ,  á_{+Àº} \} = 0âÜâA_{+ÀŒ} = 0 $$
$$ \{  á_{+[ÀŒ} ,  á_{-Àº]} \} = 0âÜâA_{-ÀŒ} = »_{+ÀŒ}A_{--} $$
(where $á=d+iA$) in terms of the ``prepotential" $A_{--}$.  But the solution of the second equation automatically follows from the first because of Lorentz invariance, when it is gauged; the prepotential appears already as a potential:  Introducing $á_{Œº}$,
$$ \{  á_{+ÀŒ} ,  á_{--} \} =  á_{-ÀŒ}âÜâA_{-ÀŒ} = »_{+ÀŒ}A_{--} $$
Pure spinors are also related to (coset) Lorentz coordinates.

Ü5.2. Background

We first express the covariant derivatives in manifestly SO(N) covariant form; their algebra is the obvious combination of the OSp(N|4) algebra with that of the flat-space super Yang-Mills covariant derivatives:
$$ \li{
[á_{\rm ab},á_{\rm cd}] & = -ú_{\rm [c|[a}á_{\rm b]|d]} \cr
[á_{\rm ab},á_{\rm c©}] & = -ú_{\rm c[a}á_{\rm b]©} \cr
Óá_{\rm aŒ},á_{\rm bº}Õ & = -C_{Œº}(á_{\rm ab}+Ä_{\rm ab})-ú_{\rm ab}á_{Œº} \cr
Óá_{\rm aÀŒ},á_{\rm bÀº}Õ & = -C_{ÀŒÀº}(á_{\rm ab}+ÐÄ_{\rm ab})-ú_{\rm ab}á_{ÀŒÀº} \cr
Óá_{\rm aŒ},á_{\rm bÀº}Õ & = -ú_{\rm ab}á_{ŒÀº} \cr
[á_{Œº},á_{\rm b©}] & = -C_{©(Œ}á_{\rm bº)} \cr
[á_{\rm aŒ},á_{©Àº}] & = C_{Œ©}(á_{\rm aÀº}+ÑW_{\rm aÀº}) \cr
[á_{\rm aÀŒ},á_{©Àº}] & = C_{ÀŒÀº}(á_{\rm a©}+W_{\rm a©}) \cr
[á_{Œº},á_{©À¶}] & = -C_{©(Œ}á_{º)À¶} \cr
[á_{ŒÀº},á_{©À¶}] & = C_{Œ©}(á_{ÀºÀ¶}+f_{ÀºÀ¶})+C_{ÀºÀ¶}(á_{Œ©}+f_{Œ©}) \cr
[á_{ÀŒÀº},á_{©À¶}] & = -C_{À¶(ÀŒ}á_{©Àº)} \cr
} $$
The Bianchi identities give almost identical results to those of flat space, the only difference being the appearance of a curvature term (resembling a mass term) in the conformal scalar field equation.  They imply the following field equations for N=4
$$ \li{
[á_{ŒÀº},ÑW_{\rm a}{}^{Àº}]-[Ä_{\rm ab},W^{\rm b}{}_Œ] & = 0 \cr
[á_Œ{}^{Àº},f_{À©Àº}]+\f14[Ä_{\rm ab},[á_{ŒÀ©},ÐÄ^{\rm ab}]]-ÓW^{\rm b}{}_Œ,ÑW_{\rm bÀ©}Õ & = 0 \cr
[á^{©Àº},[á_{©Àº},Ä_{\rm ab}]]-2ÓÑW_{\rm a}{}^{Àº},Ñ{W}_{\rm bÀº}Õ-·_{\rm abcd}ÓW^{\rm c©},W^{\rm d}{}_{©}Õ-4Ä_{\rm ab}-[Ä_{\rm bc},[Ä_{\rm ad},ÐÄ^{\rm cd}]] & = 0 \cr
} $$
and relations among the field strengths:
$$ \li{
[á_{\rm ab},Ä_{\rm cd}] & = -ú_{\rm [a[c|}Ä_{\rm b]|d]} \cr
[á_{\rm ab},W_{\rm c©}] & = -ú_{\rm c[a}W_{\rm b]©} \cr
[á_{\rm aº},Ä_{\rm cd}] & = -[á_{\rm cº},Ä_{\rm ad}] \cr
[á_{\rm aÀº},Ä_{\rm cd}] & = ú_{\rm a[c}Ñ{W}_{\rm d]Àº} \cr
Óá_{\rm aŒ},W_{\rm bº}Õ & = -ú_{\rm ab}f_{Œº} 
	+C_{Œº}(ü[Ä_{\rm ac},ÐÄ_{\rm b}{}^{\rm c}]-ÐÄ_{\rm ab})\cr 
Óá_{\rm aŒ},ÑW_{\rm bÀº}Õ & = -[á_{ŒÀº},Ä_{\rm ab}] \cr
[á_{\rm aÀŒ},f_{º©}] & = \f12[á_{(ºÀŒ},W_{\rm a©)}] \cr
[á_{\rm aÀŒ},f_{ÀºÀ©}] & = C_{ÀŒ(Àº}ÑW_{\rm aÀ©)}-\f12C_{ÀŒ(Àº}[á^¶{}_{À©)},W_{\rm a¶}] \cr
} $$
$$ Ä_{\rm ab} = ü·_{\rm abcd}ÐÄ^{\rm cd} $$
where the last is determined only up to a phase.

Ü5.3. Projective gauge

N=4 is the only projective case where the $á_u$ algebra has field strengths, but these can be absorbed by (the gauge fields of) the SO(2) derivatives.  (A similar procedure works for the N=2 chiral case, but not for N=4 chiral.)  Examining the relations
$$ \li{ Óá_{aŒ},á_{bº}Õ & = - C_{Œº}(á_{ab}+Ä_{ab}) - ú_{ab}á_{Œº} \cr
	Óá_{a'ÀŒ},á_{b'Àº}Õ & = - C_{ÀŒÀº}(á_{a'b'}+ÐÄ_{a'b'}) - ú_{a'b'}á_{ÀŒÀº} \cr} $$
$$ Ä_{ab} = C_{ab}Ä,ââÐÄ_{a'b'} = C_{a'b'}Ä,ââ[á_u, Ä] = 0 $$
(for $á_u = (á_{ab},á_{a'b'},á_{aŒ},á_{a'ÀŒ},á_{Œº},á_{ÀŒÀº})$), we see that we can consistently impose the isotropy constraints
$$ á_{Œº} = á_{ÀŒÀº} = á_{aŒ} = á_{a'ÀŒ} = á_{ab} + Ä_{ab} = á_{a'b'} + ÐÄ_{a'b'} = 0 $$
as a closed algebra.  (This is equivalent to a redefinition of the SO(2) derivatives.)  In particular, we can choose the gauge $d_u=0$ (i.e., the above minus $d_u$ = 0).  In this gauge, there is a residual gauge invariance with $d_u Â=0$; i.e., the gauge parameter is projective.  At that point we can work exclusively in terms of $á_w$. 

Some interesting features of this required modification are:  (1) It involves only the SO(n)SO(N$-$n) isotropy derivatives, and hence requires the super anti-de Sitter construction.  (The analogous derivatives in flat superspace would be central charges, which would break superconformal invariance.  However, we can still use our contracted coset, since the isotropy group is unchanged.)  (2) The modifications must involve only a single field strength to avoid generation of field-strength commutator terms (and hence nonclosure) in the algebra of isotropy constraints, and hence both n and N$-$n $²2$.  This shows that chiral superspace does not exist for N=4 Yang-Mills.

Û2 6. Four-point amplitude

Ü6.1. Duality

The four-point amplitude in this theory has been shown to have a simple form in projective superspace, where coordinate/momentum duality is almost manifest [11].  This duality is the one that results from (super) Fourier transformation, whereupon coordinates (of vertices) are replaced with loop momenta, after applying momentum conservation.  (External line momenta are also expressed as differences, by interpreting paths connecting adjacent external lines as ``half-loops", with their own momenta.)  Thus graphs are replaced with (geometrically) dual graphs [19].  In string theory, introducing a (random) lattice for the worldsheet, this is recognized as T-duality [20].  The AdS${}_5ð$S${}^5$ string has been shown to have invariance under such a T-duality [21], implying that N=4 super Yang-Mills has another PSU(4|2,2) symmetry that includes the usual Lorentz and R-symmetry, but also ``translation" invariance in the loop supermomenta (of a projective or chiral superspace), and their completion to a full dual superconformal group.

A proposal for this dual superconformal invariance of the theory had already been made directly on the N=4 Yang-Mills amplitudes [22]; however, it requires the inclusion of twistor coordinates with both the coordinate and momentum spaces, and is thus not a complete duality.  The reason why the twistors were found necessary is that this formulation is based on chiral superspace, which is simplest for MHV amplitudes.  In that space the chiral field strengths are the selfdual parts of the (superfield which at $Ï=0$ is the) Yang-Mills field strength, which carries Lorentz indices.  They thus use twistors to carry these indices.  An alternative would be to introduce spin coordinates; but these do not naturally appear in chiral superspace (at least according to our projective construction).  The scalar factor of the amplitude (the purely spacetime-momentum factor) would then acquire additional denominator factors of momenta to cancel those introduced into the chiral external line factors, since the Yang-Mills field strength has higher dimension than the scalars.  At least effective actions would be expected to be more complicated in this approach, since the chirality of this field strength holds only at the linearized level.

Ü6.2. From chiral to projective

Although we do not yet give the Feynman rules in this paper, we present an alternate derivation of this amplitude in projective superspace from chiral supertwistor space, which could be generalized to known higher-point amplitudes.  We do this not to illustrate the method, which can be complicated in general (especially if we include the effort required to derive the chiral supertwistor expressions with which we begin), but to show the simplicity of the projective superspace result.  The method is to transform the chiral supertwistor into projective supertwistor space by Fourier transforming half the fermionic twistor coordinates; the result can then be put into projective supercoordinate space by the usual (projective super) Penrose transform.  The result can already be guessed by noting that the four-point amplitude is both MHV and anti-MHV:  For the tree case, the chiral and antichiral supertwistor expressions are
$$ \A_{4} = {¶^4(Ýp_{ŒÀŒ})¶^8(ݹ_{\rm aŒ})\over Ò12ÔÒ23ÔÒ34ÔÒ41Ô} ,â
	\A_{4Ѝ} = {¶^4(Ýp_{ŒÀŒ})¶^8(Ýй^{\rm a}{}_{ÀŒ})\over [12][23][34][41]} $$

\noindent (The sums are over external lines.)  Thus we'll find that the ubiquitous twistor denominator of MHV, and its complex conjugate of anti-MHV, are replaced in projective supertwistor space by their magnitude, which is directly expressible in terms of momenta (e.g., $st$ for the tree case).

We use the notation $ijkl$ to label the 4 distinct external lines.  Then the only twistor identity we need is the equality of the MHV and anti-MHV expressions for the pure-gluon amplitude:
$$ {ÒijÔ^4\over Ò12ÔÒ23ÔÒ34ÔÒ41Ô} = {[kl]^4\over [12][23][34][41]} $$
This allows us to evaluate the fermionic Fourier transform with respect to any {\it one} of the 4 twistor fermions (with respect to N=4, but all {\it four} of the external lines):
$$ Çd^4 ½_i¼e^{iн_i ½_i}¼¶^2(Â_{iŒ}н_i) = ÝÒijÔ½_k ½_l 
	= ¶^2(ÐÂ_{iÀŒ}½_i)\left( {Ò12ÔÒ23ÔÒ34ÔÒ41Ô\over [12][23][34][41]}\right)^{1/4} $$
with Einstein summation understood on identical indices.  Thus this Fourier transformation replaces the conservation $¶$-function for total $¹_Œ=Â_ŒÐ½$ with one for the corresponding $й_{ÀŒ}=ÐÂ_{ÀŒ}½$, and throws in a phase factor.  In addition to reproducing the correct relation between the above forms of the amplitude in chiral and antichiral supertwistor space, it gives the intermediate result for projective supertwistor space:
$$ \A_{4¸} = {¶^4(Ýp_{ŒÀŒ})¶^4(ݹ_{Œa'})¶^4(Ýй_{aÀŒ})\over \f14 st} $$
Note that this amplitude is missing an explicit $¶$-function for conservation of $t_{aa'}$ (which would actually be a Kronecker $¶$, because of the compactness of the R-space):  This conservation is implied by the other $¶$-functions (in twistor superspace, or on shell).  

In this form, the amplitude is already expressed directly in momentum superspace; we need only attach external line factors, which are just the (linearized) projective superfield strengths $Ä$:
$$ ö\A_{4¸} = Çd^{16}p_i¼d^{32}¹_i¼d^{16}t_i¼÷Ä(1)÷Ä(2)÷Ä(3)÷Ä(4)
	¼{¶^4(Ýp_{ŒÀŒ})¶^4(ݹ_{Œa'})¶^4(Ýй_{aÀŒ})\over \f14 st} $$
(The $Çd^{16}t_i$ should really be a sum.  Of course, the $t_i$ conjugate to $y_i$ should not be confused with the Mandelstam variable $t$.)  For comparison, in the chiral case, we need to multiply numerator and denominator by $[12][23][34][41]$ to put the amplitude into momentum space:  The denominator becomes $(st)^2$, while for the numerator factor, including external line factors, we have
$$ {1\over Ò12ÔÒ23ÔÒ34ÔÒ41Ô} = {[12][23][34][41]\over (\f14 st)^2}
	â£â{Ðf_{ÀŒ}{}^{Àº}(1)Ðf_{Àº}{}^{À©}(2)Ðf_{À©}{}^{À¶}(3)Ðf_{À¶}{}^{ÀŒ}(4)\over (\f14 st)^2} $$
$$ Üâö\A_{4} = Çd^{16}p_i¼d^{32}¹_i¼Ðf_{ÀŒ}{}^{Àº}(1)Ðf_{Àº}{}^{À©}(2)Ðf_{À©}{}^{À¶}(3)Ðf_{À¶}{}^{ÀŒ}(4)
	¼{¶^4(Ýp_{ŒÀŒ})¶^8(ݹ_{\rm aŒ})\over (\f14 st)^2} $$
Note that there is no direct analog for the chiral supertwistor scalar wave function in momentum (or coordinate) superspace, unlike the projective case:  It is a (super)helicity amplitude, and does not directly covariantize (except by mutliplying by twistors to get tensors, as above).  This is a consequence of the fact that the only scalar superfield strength is projective, not chiral (even in the linear approximation).

We can easily Fourier transform the projective amplitude back to coordinate superspace:  The $x$ dependence is as usual, the $Ï$ dependence is the local product, and the $y$ dependence evaluates at $y=0$:
$$ö \A_{4¸} = Çd^{16}x_i¼d^8 ϼÄ(x_1,Ï,0)Ä(x_2,Ï,0)Ä(x_3,Ï,0)Ä(x_4,Ï,0)¼
	{¶^4(x_1-x_2+x_3-x_4)\over x_{12}^2 x_{23}^2} $$

Û0 7. Prospects

These results can be generalized to other N:  For example, the simpler case of N=2 would be useful to compare with the known harmonic and projective formalisms.  Since in general such first-quantization describes superspin 0, the ``smallest" supermultiplet (unless additional spin variables are included), N=2 would describe a scalar multiplet, which could also be coupled to external Yang-Mills.  The R-space in that case is SO(2), corresponding to the identification of the usual projective R-space with the unit circle.  Similarly, N=8 would describe (gauged) supergravity.  All cases N$²$8 could be coupled to external (gauged) supergravity; the formalism suggests that the tangent space for this supergravity would be OSp(n|2)OSp(N$-$n|2), rather than the purely bosonic (Lorentz and R-symmetry) tangent spaces that have been used so far.

These results can also be generalized to some other dimensions; we have at least
$$ \vbox{\halign{ D=#:â \hfil & \hfil $\rm\displaystyle#$ \hfil 
	\vrule width 0em height 2.3em \cr
5 & {(P)SU(N|2,2) \over OSp(N|4)} \cr
4 & {OSp(N|4) \over OSp(n|2)OSp(N-n|2)} \cr
3 &  {OSp(üN|2)^2 \over OSp(üN|2)} \cr
2 &  {OSp(N|2) \over U(üN|1)} \cr}} $$
The method for solving the constraints is similar, and correctly produces (at least the free) supersymmetric theories in those dimensions. 

Another problem is whether superconformal invariance can be made manifest.  A better understanding of projective lightcone limits might do that.  This might also shed some light on the relationship between the harmonic and projective approaches:  For example, for the N=2 case, we see the R-symmetry part of the coset space in D=5 is SU(2)/SO(2), as in the usual N=2 harmonic superspace, while the coset space in D=4 is just SO(2), so the sphere reduces to the circle.

In principle, superconformal invariance could be made manifest by using an action with the full superconformal set of constraints:
$$ d_{(A}{}^{(A'}d_{B]}{}^{B']} = 0 $$
However, this set is highly reducible:  In particular, it reduces to our anti-de Sitter ones
$$ d_A{}^{A'}d_B{}^{B'}ú_{B'A'},â d_A{}^{A'}d_B{}^{B'}ú^{BA} $$
(with or without contraction), and the ghosts are much messier [23].  Alternatively, one could consider the flat-space limit (R$£¥$) of the anti-de Sitter constraints,
$$ d_A{}^{ÀŒ}d_B{}^{Àº}ÐC_{ÀºÀŒ},â d_Œ{}^{A'}d_º{}^{B'}C^{ºŒ} $$
but this loses some necessary constraints:  $t_{aa'}$ drops out altogether.  These different sets of constraints can be considered as related to partial gauge fixing of the corresponding Lagrange multipliers.

It may be possible to find at least some of the superconformal invariance through transformations of the Lagrange multipliers.  For example, the usual action for a scalar particle, $ÇgÀx{}^2$, is conformal through transformation of $g$, and an (A)dS metric can be obtained by redefining $g$ by the appropriate Weyl scale factor.

An obvious topic is the gauge-invariant field theory action for N=4 Yang-Mills, and its second-quantization.  It should be noted that the supergraph rules will not be manifestly superconformal:  The second-quantized gauge-fixing term for Yang-Mills breaks conformal invariance, and first-quantization requires gauge-fixing the worldline metric, which also breaks conformal invariance.  However, it should be possible to preserve some useful affine subalgebra.

Alternatively, by finishing the treatment of first-quantization in an external N=4 super Yang-Mills background, it should be possible to define vertex operators that allow supergraph calculations directly in a first-quantized approach, in analogy to string theory.  It may then be possible to reproduce many of the results of the gauge/string correspondence without requiring the full string machinery.  For example, properties such as N=4 superconformal symmetry, or its ``dual", may be sufficient. 

Unique to the case N=4, the numbers of commuting and anticommuting coordinates cancel (at each ghost level).  This suggests that  potential zero-mode problems (and their resulting picture-changing or equivalent vacuum problems) could be directly canceled, after an appropriate (worldline-infrared) regularization.

Even if these first-quantized methods prove useful for deriving expressions for S-matrices, a more important question is whether it can be helpful in calculating anything relevant to confinement.  In this regard, a random lattice approach to the string would suggest that this first-quantized action for the N=4 superparticle might lead to a first-quantized action for a 4D N=4 superstring.

ÜAcknowledgments

This work is supported in part by National Science Foundation Grant No.¼PHY-0653342.  M.H. acknowledges CNYITP at Stony Brook for their hospitality where part of this work was done, and she also thanks Joaquim Gomis and Kiyoshi Kamimura for fruitful discussions.  M.H. is supported by the Grant-in-Aid for Scientific Research No.¼18540287.

\refs

£1 
  F.A. Berends and W.T. Giele,
  ÓNucl.\ Phys.\  B Õ{\bf 306} (1988) 759;
  \\
  F. Cachazo, P. Svr×cek and E. Witten,
  ÓJHEP Õ{\bf 0409} (2004) 006
  \xxxlink{hep-th/0403047};
  \\
  R. Britto, F. Cachazo and B. Feng,
  ÓNucl.\ Phys.\  B Õ{\bf 715} (2005) 499\\
  \xxxlink{hep-th/0412308};
  \\
  R. Britto, F. Cachazo, B. Feng and E. Witten,
  ÓPhys.\ Rev.\ Lett.\  Õ{\bf 94} (2005) 181602
  \xxxlink{hep-th/0501052};
  \\
  J.M. Drummond, J. Henn, G.P. Korchemsky and E. Sokatchev,
  \xxxlink{arXiv:0808.0491} [hep-th];
  \\
  J.M. Drummond and J.M. Henn,
  \xxxlink{0808.2475} [hep-th].
  
£2 
  Z. Bern, L.J. Dixon, D.C. Dunbar and D.A. Kosower,
  ÓNucl.\ Phys.\  B Õ{\bf 425} (1994) 217
  \xxxlink{hep-ph/9403226},
  ÓNucl.\ Phys.\  B Õ{\bf 435} (1995) 59
  \xxxlink{hep-ph/9409265};
  \\
  Z. Bern, J.S. Rozowsky and B. Yan,
  ÓPhys.\ Lett.\  B Õ{\bf 401} (1997) 273\\
  \xxxlink{hep-ph/9702424};
  \\
  Z. Bern, V. Del Duca, L.J. Dixon and D.A. Kosower,
  ÓPhys.\ Rev.\  D Õ{\bf 71} (2005) 045006
  \xxxlink{hep-th/0410224};
  \\
  R. Britto, F. Cachazo and B. Feng,
  ÓNucl.\ Phys.\  B Õ{\bf 725} (2005) 275\\
  \xxxlink{hep-th/0412103};
  \\
  Z. Bern, L.J. Dixon and D.A. Kosower,
  ÓPhys.\ Rev.\  D Õ{\bf 72} (2005) 045014
  \xxxlink{hep-th/0412210};
  \\
  Z. Bern, L.J. Dixon, D.A. Kosower, R. Roiban, M. Spradlin, C. Vergu and A. Volovich,
  ÓPhys.\ Rev.\  D Õ{\bf 78} (2008) 045007
  \xxxlink{0803.1465} [hep-th].

£3 
  L.F. Alday and J.M. Maldacena,
  ÓJHEP Õ{\bf 0706} (2007) 064
  \xxxlink{0705.0303} [hep-th];
  \\
  J.M. Drummond, G.P. Korchemsky and E. Sokatchev,
  ÓNucl.\ Phys.\  B Õ{\bf 795} (2008) 385
  \xxxlink{0707.0243} [hep-th];
  \\
  A. Brandhuber, P. Heslop and G. Travaglini,
  ÓNucl.\ Phys.\  B Õ{\bf 794} (2008) 231\\
  \xxxlink{0707.1153} [hep-th];
  \\
  J.M. Drummond, J. Henn, G.P. Korchemsky and E. Sokatchev,
  \xxxlink{0803.1466} [hep-th].

£4 
  N. Berkovits,
  ÓJHEP Õ{\bf 0004} (2000) 018
  \xxxlink{hep-th/0001035};
  \\
  N. Berkovits and B.C. Vallilo,
  ÓJHEP Õ{\bf 0007} (2000) 015
  \xxxlink{hep-th/0004171};
  \\
  N. Berkovits,
  ÓJHEP Õ{\bf 0409} (2004) 047
  \xxxlink{hep-th/0406055},
  ÓJHEP Õ{\bf 0601} (2006) 005
  \xxxlink{hep-th/0503197},
  ÓJHEP Õ{\bf 0510} (2005) 089
  \xxxlink{hep-th/0509120};
  \\
  N. Berkovits and C.R. Mafra,
  ÓJHEP Õ{\bf 0611} (2006) 079
  \xxxlink{hep-th/0607187};
  \\
  N. Berkovits and N. Nekrasov,
  ÓJHEP Õ{\bf 0612} (2006) 029
  \xxxlink{hep-th/0609012}.

£5 
  W. Siegel,
  ÓInt.\ J.\ Mod.\ Phys.\  A Õ{\bf 4} (1989) 1827;
  \\
  A. Mikovic, M. Ro×cek, W. Siegel, P. van Nieuwenhuizen, J. Yamron and A.E. van de Ven,
  ÓPhys.\ Lett.\  B Õ{\bf 235} (1990) 106;
  \\
  K. Lee and W. Siegel,
 ÓJHEP Õ{\bf 0508} (2005) 102
  \xxxlink{hep-th/0506198},
 ÓJHEP Õ{\bf 0606} (2006) 046
  \xxxlink{hep-th/0603218}.

£6 
  G.G. Hartwell and P.S. Howe,
  ÓInt.\ J.\ Mod.\ Phys.\  A Õ{\bf 10} (1995) 3901\\
  \xxxlink{hep-th/9412147},
  ÓClass.\ Quant.\ Grav.\  Õ{\bf 12} (1995) 1823;
  \\
  P. Heslop and P.S. Howe,
  ÓClass.\ Quant.\ Grav.\  Õ{\bf 17} (2000) 3743
  \xxxlink{hep-th/0005135};
  \\
  P.J. Heslop,
  ÓClass.\ Quant.\ Grav.\  Õ{\bf 19} (2002) 303
  \xxxlink{hep-th/0108235}.

£7 
  M. Hatsuda and W. Siegel,
  ÓPhys.\ Rev.\  D Õ{\bf 67} (2003) 066005
  \xxxlink{hep-th/0211184};
  ÓPhys.\ Rev.\  D Õ{\bf 77} (2008) 065017
  \xxxlink{0709.4605} [hep-th].

£8 
  A. Karlhede, U. Lindstr¬om and M. Ro×cek,
  ÓPhys.\ Lett.\  B Õ{\bf 147} (1984) 297;
  \\
  U. Lindstr¬om and M. Ro×cek,
  ÓCommun.\ Math.\ Phys.\  Õ{\bf 115} (1988) 21,
  {\bf 128} (1990) 191;
  \\
  F. Gonzalez-Rey, M. Ro×cek, S. Wiles, U. Lindstr¬om and R. von Unge,
  ÓNucl.\ Phys.\  B Õ{\bf 516} (1998) 426
  \xxxlink{hep-th/9710250};
  \\
  F. Gonzalez-Rey and R. von Unge,
  ÓNucl.\ Phys.\  B Õ{\bf 516} (1998) 449\\
  \xxxlink{hep-th/9711135};
  \\
  F. Gonzalez-Rey,
  \xxxlink{hep-th/9712128};
  \\
  F. Gonzalez-Rey and M. Ro×cek,
  ÓPhys.\ Lett.\  B Õ{\bf 434} (1998) 303
  \xxxlink{hep-th/9804010}.

£9 
  A. Galperin, E. Ivanov, S. Kalitsyn, V. Ogievetsky and E. Sokatchev,
  ÓPhys.\ Lett.\  B Õ{\bf 151} (1985) 215;
  \\
  F. Delduc and J. McCabe,
  ÓClass.\ Quant.\ Grav.\  Õ{\bf 6} (1989) 233.

£10 
  I.L. Buchbinder, O. Lechtenfeld and I.B. Samsonov,
   ÓNucl.\ Phys.\  B Õ{\bf 802} (2008) 208 
  \xxxlink{0804.3063} [hep-th].
£11 
  R. Kallosh,
 \xxxlink{0711.2108} [hep-th].

£12 
  M. Hatsuda and K. Kamimura,
  ÓNucl.\ Phys.\  B Õ{\bf 611} (2001) 77
  \xxxlink{hep-th/0106202}.

£13 
  W. Siegel,
  ÓPhys.\ Rev.\  D Õ{\bf 52} (1995) 1042
  \xxxlink{hep-th/9412011}.

£14 
  N. Berkovits,
  ÓJHEP Õ{\bf 0708} (2007) 011
  \xxxlink{hep-th/0703282}.

£15 
    W. Siegel,
  ÓPhys.\ Rev.\  D Õ{\bf 47} (1993) 2504 
\xxxlink{hep-th/9207043}.

£16 
  W. Siegel,
  ÓPhys.\ Lett.\  B Õ{\bf 203} (1988) 79.

£17 
  S. Deser,
  ÓClass.\ Quant.\ Grav.\  Õ{\bf 4} (1987) L99.

£18 
  A. Galperin, E. Ivanov, S. Kalitsyn, V. Ogievetsky and E. Sokatchev,
  ÓClass.\ Quant.\ Grav.\  Õ{\bf 1} (1984) 469;
  \\
  A. Galperin, E. Ivanov, V. Ogievetsky and E. Sokatchev,
  ÓJETP Lett.\  Õ{\bf 40} (1984) 912
  [ÓPisma Zh.\ Eksp.\ Teor.\ Fiz.\  Õ{\bf 40} (1984) 155],
  ÓClass.\ Quant.\ Grav.\  Õ{\bf 2} (1985) 601,
  617.

£19 
  F. David and R. Hong Tuan,
  ÓPhys.\ Lett.\  B Õ{\bf 158} (1985) 435.

£20 
  W. Siegel,
  ÓPhys.\ Lett.\  B Õ{\bf 252} (1990) 558.

£21 
  N. Berkovits and J. Maldacena,
  ÓJHEP Õ{\bf 0809} (2008) 062
  \xxxlink{0807.3196} [hep-th];
  \\
  N. Beisert, R. Ricci, A.A. Tseytlin and M. Wolf,
  \xxxlink{0807.3228} [hep-th].

£22 
  J.M. Drummond, J. Henn, G.P. Korchemsky and E. Sokatchev,
  \xxxlink{0807.1095} [hep-th].

£23 
  U. Lindstr¬om, M. Ro×cek, W. Siegel, P. van Nieuwenhuizen and A.E. van de Ven,
  ÓJ.\ Math.\ Phys.\  Õ{\bf 31} (1990) 1761.

\bye